\documentclass[11pt,a4paper]{article}
\usepackage[dvips]{graphics}
\makeatletter \oddsidemargin -.1in \evensidemargin -.1in
\newcommand{\singlespacing}{\let\CS=\@currsize\renewcommand{\baselinestretch}{1}\tiny\CS}
\newcommand{\doublespacing}{\let\CS=\@currsize\renewcommand{\baselinestretch}{1.35}\tiny\CS}
\usepackage{graphics}
\usepackage{amsmath}
\usepackage{amssymb}
\usepackage{float, epsfig, floatflt, here}
\usepackage{cite}
\usepackage{longtable}
\usepackage{graphicx}
\usepackage{epstopdf}
\usepackage{pdfpages}
\usepackage[utf8]{inputenc}
\usepackage[T1]{fontenc}
\usepackage{qcircuit}
\usepackage{caption}
\usepackage{subcaption}
\title {\textbf{\Large Simultaneous Asymmetric Quantum Remote State Preparation Scheme in Noisy Environments}}
\author{ \textbf{ Binayak S. Choudhury}$^{1}$ \thanks {e-mail : binayak@math.iiests.ac.in, Fax: +91-33-26682916},\textbf{ Manoj Kumar Mandal}$^{1}$ \thanks{ e-mail: manojmandaliiest@gmail.com, Tel: +91-700395967},
\\ \textbf{Soumen Samanta}$^{1}$ \thanks{corresponding author e-mail : s.samanta.math@gmail.com, Tel: +91-9804842707}, \textbf{Biswanath Dolai}$^{1,2}$ \thanks{e-mail : biswanathbmm@gmail.com, Tel: +91-9547360567}
\\   $^{1}$ Department of Mathematics,
 Indian Institute of Engineering Science and Technology,\\ Shibpur
 B. Garden, Howrah - 711103,
 West Bengal, India\\
$^{2}$ Department of Physics,
 Bajkul Milani Mahavidyalaya, Bajkul, \\Purba Medinipur-721655,
 West Bengal, India\\}
 
\begin{document}
\date{}
\maketitle

\noindent{\textbf{Abstract}} In this paper we discuss a quantum multi-tasking protocol for preparation of known one-qubit and two-qubit states respectively in two different locations. The ideal remote state preparation protocol is discussed in the first place in which a five-qubit entangled state is utilized. The design for the preparation of such entanglement is also presented and run on IBM quantum composer platform. The effects of three types of noises on the protocol are discussed and in all these cases the reduced fidelities of the process are calculated. The variations of these fidelities with respect to different parameters are analysed. \\


\section{\textbf{ Introduction}}
Quantum teleportation (QT) introduced by Bennett et al. \cite{bennett} in 1993 plays an important role in the history of quantum communication and information theory. It is a technique for transfer of unknown quantum information from one location to a distant location through a previously shared entanglement between the sender and the receiver. Since then, a series of modified quantum teleportation protocols such as controlled teleportation \cite{kb, zm, z, gyl, lln, hbzy, cllyz, sm}, bi-directional teleportation \cite{ln, css, zxl, zz, jzl, ka, cbs}, probabilistic teleportation \cite{ap, yw, yy, llc, mzcw, jpyp}, cyclic teleportation \cite{cdlw, zlz, wbz, csb, zl, vym}, multi-hop teleportation \cite{zyxz, cbss, zyz, whs, f} and so on were developed  in course of the last three decades which are applicable to a very wide variety of quantum states.\\

\noindent There is another class of quantum communication scheme somewhat similar to quantum teleportation protocols in the sense that its aim is also to create a quantum state at a distant location but the prime difference is that the state to be remotely prepared is known to the sender. These types of quantum communication schemes are known as quantum remote state preparation (RSP) protocols which were first introduced by Pati \cite{p}. Following the work of Pati \cite{p}, several researchers developed various modified forms of quantum remote state preparation schemes such as controlled RSP \cite{zswt, zyc, s}, joint remote state preparation \cite{csbb, sy, pbty}, cyclic RSP \cite{jh, mbs, mclz}, bi-directional RSP \cite{lqsn, gl, cmsd}, conference RSP \cite{bscs}, etc.\\

\noindent However, in the process of quantum state transfer or creation of states in remote locations there will inevitably be interactions with the environment. As a result of this random coupling with the environment, the shared entangled channel is distorted which gives rise to quantum noisy channels. In a noisy channel quantum decoherence due to interaction with the environment is described by Kraus operators which are in general non-unitary and therefore irreversible. In recent times, some remote state preparation schemes through noisy environments  are discussed in \cite{zh, swq, jzxl, ji, gcxy}. The conventional measure of effectiveness in quantum communication is the fidelity between the input state and the output state. The effect of noise is the reduction in fidelity.\\

\noindent In this paper we describe a quantum communication scheme for remotely preparing a single qubit and a two-qubit quantum state respectively at the sites of two different parties. Both the states are known to the party intending to remotely create the two states. We first describe the ideal case where the task is performed through the utilization of a five-qubit cluster state in which there is no infiltration of noise. The preparation of the five-qubit entangled channel used here is also discussed. After that we consider the effect of noisy environment in our protocol. Specifically we discuss three types of noises Amplitude-damping, Phase-flip and Bit-flip noise. Due to the presence of noise the fidelity of the process which is a measure of effectiveness of the communication is decreased. We make a study of the variation of fidelity with respect to different parameters of the scheme.

\section{\textbf{Simultaneous quantum remote state preparation in ideal case }}
 Suppose there are three parties, namely, Alice, Bob and Candy who are situated at three different locations. Let us consider an arbitrary single-qubit and a two-qubit quantum state which are respectively given by\\ 
$~~~~~~~~~~~~~~~~~~~~~~|S_1\rangle=(\alpha|0\rangle+\beta|1\rangle), \hfill(1)$\\
$~~~~~~~~~~~~~~~~~~~~~~|S_2\rangle=(\gamma|00\rangle+\delta|11\rangle), \hfill(2)$\\
where the coefficients $\alpha, \beta, \gamma$ and $\delta$ meet the normalization conditions, that is, $|\alpha|^{2}+|\beta|^{2}=1$ and $|\gamma|^{2}+|\delta|^{2}=1$. \\

\noindent A five-qubit cluster-state is given by\\
$~~~~~~~~~~~~~~~~~|\Phi\rangle=\frac{1}{2}(|00000\rangle+|01011\rangle+|10100\rangle-|11111\rangle)_{12345}, \hfill(3)$\\
which acts as the quantum channel.\\

\noindent Let us consider the situation where Alice intends to remotely create the two states given in (1) and (2) in the locations of Bob and Candy respectively. Alice wants to prepare the quantum state $|S_1\rangle$ at Bob's laboratory and at the same time she wants to create the quantum state $|S_2\rangle$ at Candy's site. The sender Alice possesses the knowledge of the coefficients  $\alpha, \beta, \gamma$ and $\delta$. For this purpose Alice, Bob and Candy share a five-qubit cluster-state given in (3), in which the particles ($1, 2$) belong to Alice, particle 3 belongs to Bob and particles ($4, 5$) are in the hands of Candy. \\

\noindent For our purpose we rewrite the quantum entangled channel given in (3) as \\
$~~~~~~~~~~~~~~~~~|\Phi\rangle=\frac{1}{2}(|00000\rangle+|01011\rangle+|10100\rangle-|11111\rangle)_{a_1a_2b_1c_1c_2}. \hfill(4)$\\

\noindent Alice performs a 2-qubit measurement in the basis consisting of four linearly independent vectors
given by\\
$~~~~~~~~~~~~~~~~~~~~~~|\xi_1\rangle=(\alpha \gamma|00\rangle+ \alpha \delta |01\rangle+\beta \gamma|10\rangle-\beta \delta |11\rangle)_{a_1a_2}, $\\
$~~~~~~~~~~~~~~~~~~~~~~|\xi_2\rangle=(\alpha \delta|00\rangle- \alpha \gamma |01\rangle-\beta \delta|10\rangle-\beta \gamma |11\rangle)_{a_1a_2}, $\\ 
$~~~~~~~~~~~~~~~~~~~~~~|\xi_3\rangle=(\beta \gamma|00\rangle+ \beta \delta |01\rangle-\alpha \gamma|10\rangle+\alpha \delta |11\rangle)_{a_1a_2}, $\\
$~~~~~~~~~~~~~~~~~~~~~~|\xi_4\rangle=(\beta \delta|00\rangle- \beta \gamma |01\rangle+\alpha \delta|10\rangle+\alpha \gamma |11\rangle)_{a_1a_2}. \hfill(5) $\\

\noindent The choice of such a measurement base by Alice is possible since the coefficients 
$\alpha , \beta , \gamma , \delta$ are known to Alice.\\

\noindent We can express the above quantum channel given in (4) using the basis (5) as\\
$~~~~~~~~~~~~~~~|\Phi\rangle=\frac{1}{2}\bigg[|\xi_1\rangle_{a_1a_2}  \otimes (\alpha \gamma |000\rangle+\alpha \delta |011\rangle+\beta \gamma |100\rangle+\beta \delta |111\rangle)_{b_1c_1c_2}\\
~~~~~~~~~~~~~~~~~~~~~~~+|\xi_2\rangle_{a_1a_2} \otimes (\alpha \delta |000\rangle-\alpha \gamma |011\rangle-\beta \delta |100\rangle+\beta \gamma |111\rangle)_{b_1c_1c_2}\\
~~~~~~~~~~~~~~~~~~~~~~~+|\xi_3\rangle_{a_1a_2} \otimes (\beta \gamma |000\rangle+\beta \delta |011\rangle-\alpha \gamma |100\rangle-\alpha \delta |111\rangle)_{b_1c_1c_2}\\
~~~~~~~~~~~~~~~~~~~~~~~+|\xi_4\rangle_{a_1a_2} \otimes (\beta \delta |000\rangle-\beta \gamma |011\rangle+\alpha \delta |100\rangle-\alpha \gamma |111\rangle)_{b_1c_1c_2}\bigg]\\
~~~~~~~~~~~~~~~~~~~=\frac{1}{2}\bigg[|\xi_1\rangle_{a_1a_2}  \otimes (\alpha |0\rangle+\beta |1\rangle)_{b_1} \otimes (\gamma |00\rangle+\delta |11\rangle)_{c_1c_2}\\
~~~~~~~~~~~~~~~~~~~~~~~+|\xi_2\rangle_{a_1a_2} \otimes (\alpha |0\rangle-\beta |1\rangle)_{b_1} \otimes (\delta |00\rangle-\gamma |11\rangle)_{c_1c_2}\\
~~~~~~~~~~~~~~~~~~~~~~~+|\xi_3\rangle_{a_1a_2} \otimes (\beta |0\rangle-\alpha |1\rangle)_{b_1} \otimes (\gamma |00\rangle+\delta |11\rangle)_{c_1c_2}\\
~~~~~~~~~~~~~~~~~~~~~~~+|\xi_4\rangle_{a_1a_2} \otimes (\beta |0\rangle+\alpha |1\rangle)_{b_1} \otimes (\delta |00\rangle-\gamma |11\rangle)_{c_1c_2}\bigg].\hfill(6)$\\

\noindent After the measurement Alice sends her outcome to Bob and Candy simultaneously using two separate classical communication channel. Receiving the measurement results from the sender Alice, the other two parties Bob and Candy make appropriate unitary operations on their respective particles to get the intended quantum states. That is the end of the protocol. The details of Alice's measurement outcomes and corresponding appropriate unitary operations are given in Table 1.\\ 

\noindent As an illustration, suppose Alice's measurement outcome is $|\xi_2\rangle$ then the unitary operator applied by Bob and Candy is $(\sigma_z)_{b_1}$ and $(\sigma_x)_{c_1} \otimes (\sigma_x\sigma_z)_{c_2}$ respectively.\\
\begin{center}
    \begin{tabular}{ |p{4cm}|p{3cm}|p{3cm}|  }
\hline
\multicolumn{3}{|c|}{Table 1: Outcome results and corresponding unitary operators} \\
\hline
Measurement outcomes of Alice ($|\xi_i$) & Bob's unitary operator ($U^i$) &Candy's unitary operator ($V^i$) \\
\hline
$|\xi_1\rangle_{a_1a_2}$ & $(I)_{b_1}$ & $(I)_{c_1}\otimes (I)_{c_2}$ \\
$|\xi_2\rangle_{a_1a_2}$ & $(\sigma_z)_{b_1}$ &$(\sigma_x)_{c_1} \otimes (\sigma_x\sigma_z)_{c_2}$ \\
$|\xi_3\rangle_{a_1a_2}$ & $(\sigma_x\sigma_z)_{b_1}$ & $(I)_{c_1}\otimes (I)_{c_2}$ \\
$|\xi_4\rangle_{a_1a_2}$ & $(\sigma_x)_{b_1}$ & $(\sigma_x)_{c_1} \otimes (\sigma_x\sigma_z)_{c_2}$\\
\hline
\end{tabular}
\end{center}

\section{Preparation of Entangled Channel}
  The circuit for generating the quantum state $|\Phi\rangle$ in (3) is given in Fig. 1. It is generated by utilizing two Hadamard gates and four CNOT gates.\\
\noindent Initially, a five-qubit state is prepared from a five-zero initial state\\
 $$|\psi_1\rangle=|0\rangle_1\otimes|0\rangle_2\otimes|0\rangle_3\otimes|0\rangle_4\otimes|0\rangle_5.$$\\
Now, first, one Hadamard gate is applied on qubit 1 and then a CNOT gate is applied with qubit 1 as a control qubit and qubit 2 as the target qubit. Then the initial state $|\psi_1\rangle$ is converted to the state \\
$$  |\psi_2\rangle=\frac{1}{\sqrt{2}}\Big(|00000\rangle+|11000\rangle\Big)_{12345.}$$\\
 Again one Hadamard gate is applied on qubit 2 then the state $|\psi_2\rangle$  becomes\\
  $$|\psi_3\rangle=\frac{1}{2}\Big(|00000\rangle+|01000\rangle+|10000\rangle-|11000\rangle\Big)_{12345.}$$\\
\noindent In the next step,  one CNOT gate is applied with qubit 1 as the control qubit and qubit  3 as the target qubit. After that  two CNOT gates are applied with qubit 2 as the control qubit for each of qubits 4 and 5 respectively as target qubits. Then the state $|\psi_3\rangle$ is transferred to\\
   $$|\Phi\rangle=\frac{1}{2}\Big(|00000\rangle+|01011\rangle+|10100\rangle-|11111\rangle\Big)_{12345,}$$\\
   \noindent which is the same as (3), that is, the entanglement resource we use here.\\
   
\begin{figure}[t]
\includegraphics[width=0.9\linewidth, height=6cm]{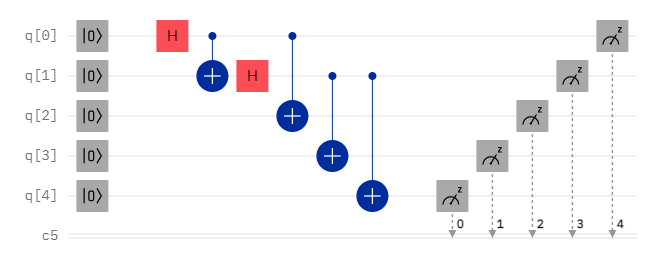} 
\caption{Quantum circuit }
\label{1}
\end{figure}

\begin{figure}[b]
\includegraphics[width=0.9\linewidth, height=5cm]{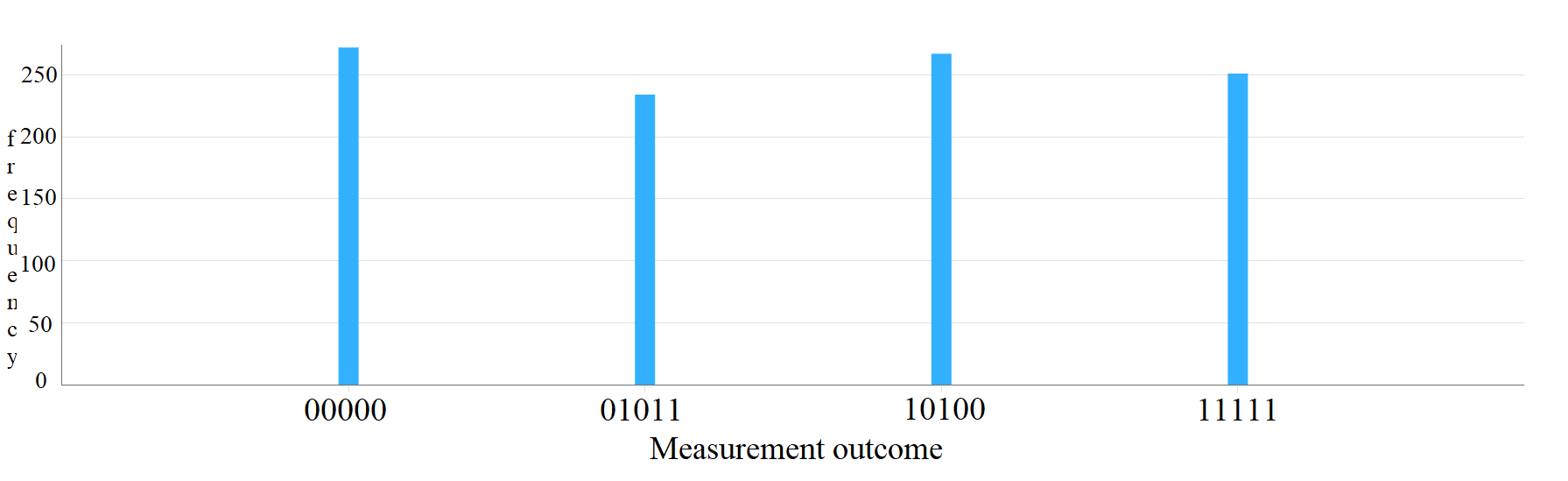} 
\caption{Quantum circuit }
\label{2}
\end{figure}

We have executed this scheme of constructing the entangled state $|\Phi\rangle$ on IBM Quantum Composer and run over $ibmq\_qasm\_simulator$ of 32 qubits. The explicit circuit and output results are given in Fig.\ref{1} and Fig.\ref{2} respectively.\\
 
\section{\textbf{Effect of noise on the proposed simultaneous remote state preparation protocol}}
In the practical field of quantum communication theory, mainly in the cases of quantum teleportation and quantum remote state preparation schemes, quantum noise is an inevitable phenomenon and it is necessary to take noise into account in our present simultaneous remote state preparation scheme. Here, we discuss the effects of three types of noises on our protocol which is a perfect communication scheme in the ideal situation where noise is neglected. We consider three different types of noises, namely, Amplitude-damping, Phase-flip and Bit-flip noisy environment.\\

\noindent Suppose Alice creates the 5-qubit entangled channel in her laboratory, keeps her own qubits $(a_1, a_2)$ with her and transmits the qubit $b_1$ to Bob and the qubits $(c_1, c_2)$ to Candy through noisy quantum channels. Thus only the qubits $b_1, c_1, c_2$ are affected by the noise as these qubits are distributed through the noisy environment and the qubits $a_1, a_2$ are not influenced by any environmental noise since these are not transferred via noisy environment.\\

\noindent In section 2, the proposed scheme is described by using a pure 5-qubit entangled state $|\Phi\rangle$ given in equation (4). The corresponding density matrix can be written as $\rho = |\Phi\rangle \langle \Phi|$. However, after passing through noisy environments the corresponding density matrix $\rho$ can be rewritten as
$$\epsilon(\rho)=\sum_{l,m,n} \big(I \otimes I \otimes X_l \otimes X_m \otimes X_n \big)~ \rho ~\big(I \otimes I \otimes X_l \otimes X_m \otimes X_n  \big)^{\dagger}~~~~~~~~~~~~ (7)$$ \\
where $X_l$ s, $X_m $s, $X_n s$ are the Kraus operators satisfy $\sum_j X_j^{\dagger} X_j=I$ and $`\dagger'$ denotes conjugate transpose of a matrix.\\

\noindent After that all the parties do the same job as in an ideal situation when we do not consider the effect of environmental noise in the said protocol. The intended remote state preparations are thereby completed.\\

\noindent The final output state $\rho_i^{out}$, where $i \in \{1,2,3,4\}$ can be calculated as \\
$~~~~~~~~~~~~~~~~~~~~~~~~\rho_i^{out}= Tr_{a_1a_2}\{M_i[\epsilon(\rho)]M_i^{\dagger}\}, \hfill (8)$\\
where partial trace is taken over the qubit pair $(a_1, a_2)$ and $M_i$ is given by\\
$~~~~~~~~~~~~~~~~~~M_i=\{I_{a_1a_2} \otimes (U^i)_{b_1} \otimes (V^i)_{c_1c_2}\}\{|\xi_i\rangle_{a_1a_2} \langle \xi_i|\otimes I_{b_1c_1c_2}\} \hfill (9)$\\
in which $U^{i}$ s and $V^{i}$ s are unitary operators given in Table 1.\\

\noindent The fidelity corresponding to the output state $\rho_i^{out}$ can be computed as \\
$~~~~~~~~~~~~~~~~~F=\langle \Psi| \rho_i^{out} |\Psi\rangle, \hfill (10)$\\
where $|\Psi\rangle$ represents the ideal output state which is the same as the input state. In our proposed scheme, it is given by\\
$~~~~~~~~~~~~~|\Psi\rangle=(\alpha |0\rangle+\beta |1\rangle)_{b_1} \otimes (\gamma |00\rangle+\delta |11\rangle)_{c_1c_2}$.\\

\noindent In the following we separately discuss the effects of three types of noises on our otherwise ideally noise-free protocol.

\subsection{\textbf{ Amplitude-damping noisy environment  }}
The Kraus operator of an amplitude-damped noisy channel is expressed as:\\
$$X_0=\left( 
\begin{array}{rr}
1 & 0 \\ 
0 & \sqrt{1-\lambda}%
\end{array}%
\right)~~~~~~~~~~X_1=\left( 
\begin{array}{rr}
0 & \sqrt{\lambda} \\ 
0 & 0%
\end{array}%
\right),$$
where $\lambda \in [0,1]$ indicates the decoherence rate of amplitude-damping noise.\\

\noindent Due to the random interaction with environment the energy of the quantum state will be dissipated. After the qubits $b_1, c_1, c_2$ are transmitted through the amplitude-damping noisy environment, the density matrix of the transformed quantum channel according to the formula (7) is given as (ignoring the constant factor)\\
$\epsilon^{AD}(\rho)=\bigg[ \bigg(|00000\rangle +(1-\lambda) |01011\rangle +\sqrt{1-\lambda} |10100\rangle -(1-\lambda)^{3/2} |11111\rangle \bigg) \times \bigg(\langle00000|\\~~~~~~~~~~ +(1-\lambda) \langle01011| +\sqrt{1-\lambda} \langle10100| -(1-\lambda)^{3/2} \langle11111| \bigg) \bigg] + \bigg[ \bigg(\sqrt{\lambda (1-\lambda)} |01010\rangle \\~~~~~~~~~~ -(1-\lambda) \sqrt{\lambda} |11110\rangle \bigg) \times \bigg(\sqrt{\lambda (1-\lambda)} \langle01010| -(1-\lambda) \sqrt{\lambda} \langle11110| \bigg) \bigg] + \bigg[ \bigg(\sqrt{\lambda (1-\lambda)} |01001\rangle \\~~~~~~~~~~ -(1-\lambda) \sqrt{\lambda} |11101\rangle \bigg) \times \bigg(\sqrt{\lambda (1-\lambda)} \langle01001| -(1-\lambda) \sqrt{\lambda} \langle11101| \bigg) \bigg] +\bigg[ \bigg(\lambda |01000\rangle \\~~~~~~~~~~ - \lambda \sqrt{1-\lambda} |11100\rangle \bigg) \times \bigg(\lambda \langle01000| - \lambda \sqrt{1-\lambda} \langle11100| \bigg) \bigg] + \bigg[ \bigg( \sqrt{\lambda} |10000\rangle -(1-\lambda) \sqrt{\lambda} |11011\rangle \bigg) \\~~~~~~~~~~ \times \bigg(\sqrt{\lambda} \langle10000| -(1-\lambda) \sqrt{\lambda} \langle11011|  \bigg) \bigg] +(1-\lambda) \lambda^2 \bigg(|11010\rangle \langle11010| +|11001\rangle \langle11001|\bigg) \\ ~~~~~~~~~~+ \lambda^3 \bigg(|11000\rangle \langle11000|\bigg)_{a_1a_2b_1c_1c_2}. \hfill (11)$\\

\noindent Now Alice performs her measurement with the measuring basis given in (5) and sends her measurement results to Bob and Candy classically. In order to get the desired state, Bob and Candy make appropriate unitary operations on their respective qubits.\\

\noindent As an illustration, suppose Alice's measurement outcome is $|\xi_2\rangle$.  Therefore the density matrix of the final output state corresponding to the Alice's measurement outcome $|\xi_2\rangle$ according to the formula (8) is given by\\
$\rho_{2}^{AD}=\bigg[ \bigg(\alpha \delta|011\rangle +(1-\lambda) \alpha \gamma |000\rangle +\sqrt{1-\lambda} \beta \delta |111\rangle +(1-\lambda)^{3/2} \beta \gamma |100\rangle \bigg) \times \bigg(\alpha \delta\langle011|\\~~~~~~~~ +(1-\lambda) \alpha \gamma\langle000| +\sqrt{1-\lambda} \beta \delta \langle111| +(1-\lambda)^{3/2} \beta \gamma\langle100| \bigg) \bigg] + \bigg[ \bigg(-\sqrt{\lambda (1-\lambda)} \alpha \gamma|001\rangle \\~~~~~~~~ -(1-\lambda) \sqrt{\lambda} \beta \gamma |101\rangle \bigg) \times \bigg(-\sqrt{\lambda (1-\lambda)} \alpha \gamma \langle001| -(1-\lambda) \sqrt{\lambda} \beta \gamma \langle101| \bigg) \bigg] \\~~~~~~~~+ \bigg[ \bigg(\sqrt{\lambda (1-\lambda)} \alpha \gamma |010\rangle  +(1-\lambda) \sqrt{\lambda} \beta \gamma|110\rangle \bigg) \times \bigg(\sqrt{\lambda (1-\lambda)} \alpha \gamma\langle010|+(1-\lambda) \sqrt{\lambda}\\~~~~~~~ \beta \gamma \langle110| \bigg) \bigg] +\bigg[ \bigg(-\lambda \alpha \gamma |011\rangle  -\lambda \sqrt{1-\lambda} \beta \gamma|111\rangle \bigg) \times \bigg(-\lambda \alpha \gamma \langle011| - \lambda \sqrt{1-\lambda} \beta \gamma \langle111| \bigg) \bigg]\\~~~~~~~~~~ + \bigg[ \bigg( -\sqrt{\lambda} \beta \delta |011\rangle -(1-\lambda) \sqrt{\lambda} \beta \gamma |000\rangle \bigg)  \times \bigg(-\sqrt{\lambda} \beta \delta \langle011| -(1-\lambda) \sqrt{\lambda} \beta \gamma\langle000|  \bigg) \bigg] \\~~~~~~~~~~+(1-\lambda) \lambda^2 \beta \gamma\bigg(|001\rangle \langle001| +|010\rangle \langle010|\bigg) + \lambda^3 \beta \gamma \bigg(|011\rangle \langle011|\bigg)_{b_1c_1c_2}. \hfill (12)$\\

\noindent The fidelity of the final output state can be obtained according to the formula (10) as\\
$~~~~~~~F_1=\langle \Psi|\rho_2^{AD}|\Psi\rangle\\
~~~~~~~~~=\bigg(\alpha^2 \delta^2 +(1-\lambda) \alpha^2 \gamma^2 + \sqrt{1-\lambda} \beta^2 \delta^2 + (1-\lambda)^{3/2} \beta^2 \gamma^2 \bigg)^2 + \bigg(\lambda \alpha^2 \gamma \delta + \lambda \sqrt{1-\lambda}\beta^2 \gamma \delta \bigg)^2\\
~~~~~~~~~~~~~~~+\bigg(\sqrt{\lambda} \alpha \beta \delta^2 + (1-\lambda) \sqrt{\lambda} \alpha \beta \gamma^2 \bigg)^2 +\lambda^3 \alpha^2 \beta^2 \gamma^2 \delta^2$.
\begin{figure}[t]
\begin{subfigure}{0.5\textwidth}
\includegraphics[width=0.9\linewidth, height=5cm]{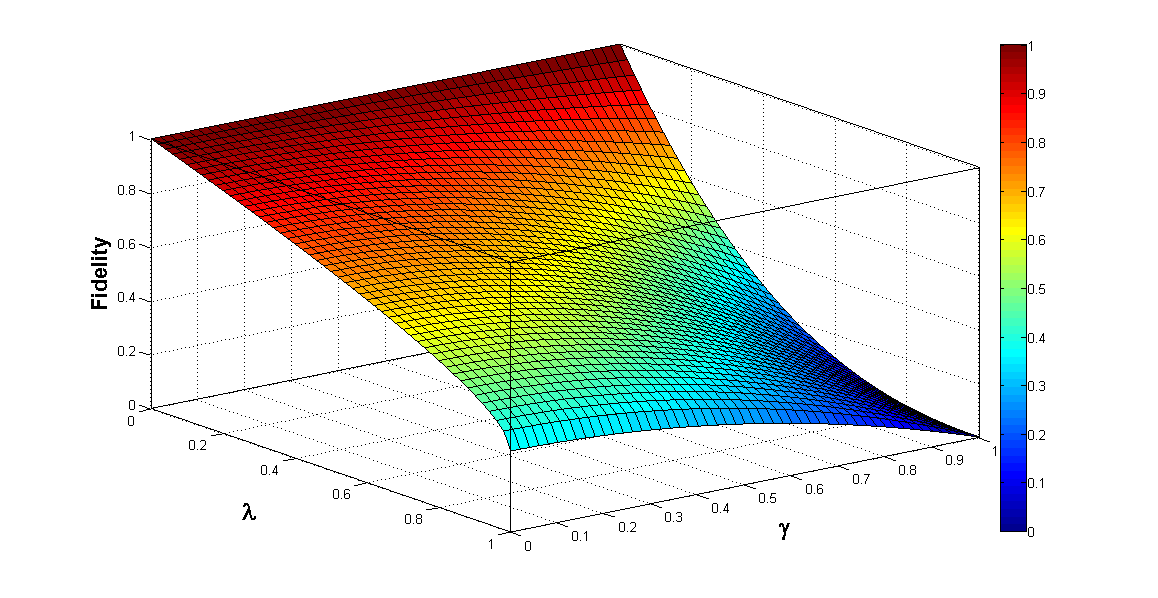} 
\caption{ }
\label{}
\end{subfigure}
\begin{subfigure}{0.5\textwidth}
\includegraphics[width=0.9\linewidth, height=5cm]{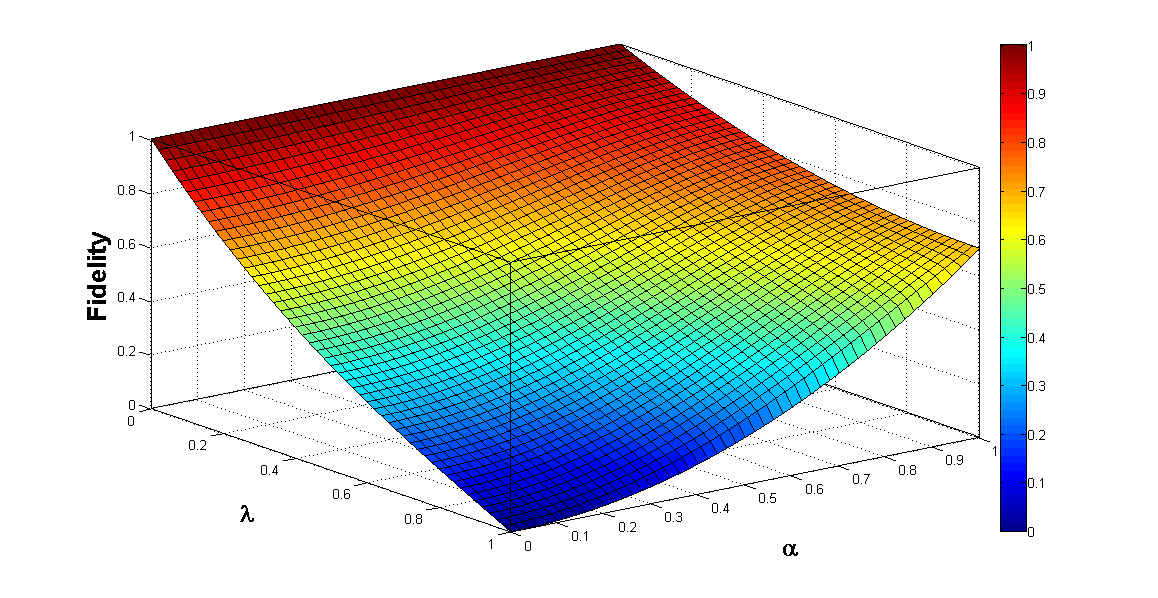} 
\caption{ }
\label{}
\end{subfigure}
\begin{subfigure}{0.5\textwidth}
\includegraphics[width=0.9\linewidth, height=5cm]{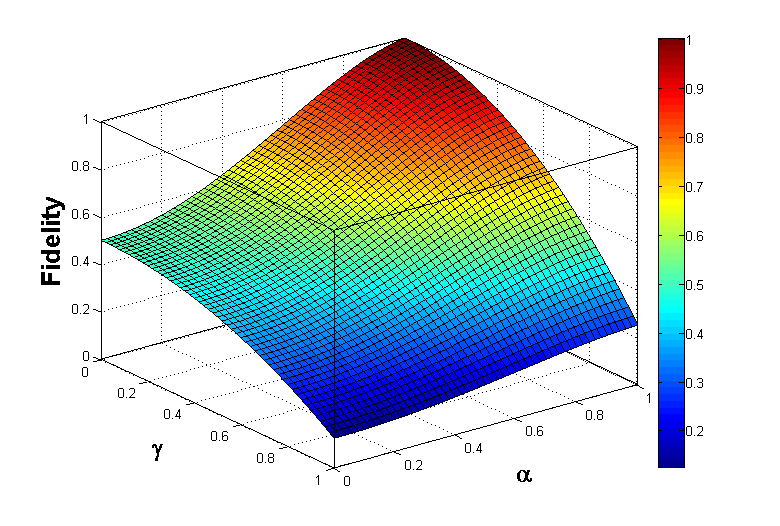} 
\caption{ }
\label{}
\end{subfigure}
\begin{subfigure}{0.5\textwidth}
\includegraphics[width=0.9\linewidth, height=5cm]{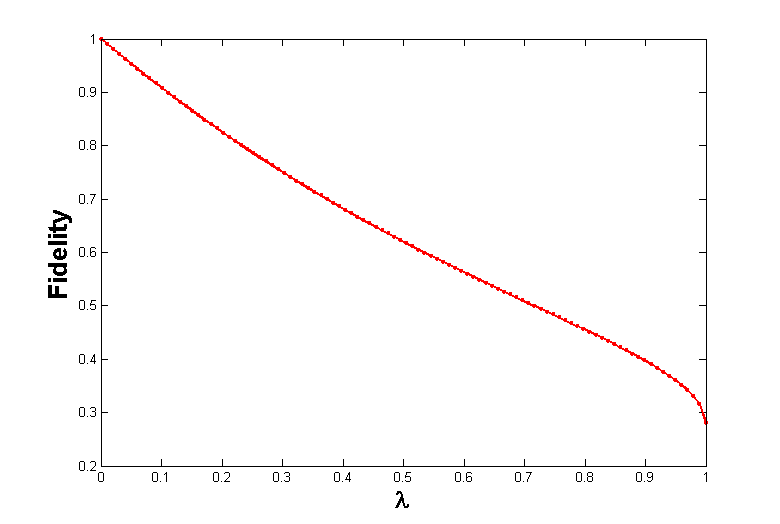} 
\caption{ }
\label{}
\end{subfigure}
\caption{(a): Variation of fidelity $F_1$  with $\gamma$ and $\lambda$ when $ |S_1\rangle=(\sqrt{0.3}|0\rangle+\sqrt{0.7}|1\rangle)$ (b): Variation of fidelity $F_1$  with $\alpha$ and $\lambda$ when  $|S_2\rangle=(\sqrt{0.3}|00\rangle+\sqrt{0.7}|11\rangle)$ (c): Variation of fidelity $F_1$  with $\alpha$ and  $\gamma$ when  $\nu=0.5$ (d) Variation of fidelity $F_1$  with $\lambda$ when $ |S_1\rangle=(\sqrt{0.4}|0\rangle+\sqrt{0.6}|1\rangle)$ and $|S_2\rangle=(\sqrt{0.3}|00\rangle+\sqrt{0.7}|01\rangle)$. }
\label{fig:1}
\end{figure}
\noindent \subsection{\textbf{ Phase-flip noisy environment  }}
The Kraus operators for the phase-flip noisy environment are given below:\\
$$X_0=\sqrt{1-\mu}\left( 
\begin{array}{rr}
1 & 0 \\ 
0 & 1%
\end{array}%
\right)~~~~~~~~~~X_1=\sqrt{\mu}\left( 
\begin{array}{rr}
1 & 0\\ 
0 & -1%
\end{array}%
\right),$$
where $\mu \in [0,1]$ defines the noise parameter in phase-flip noisy environment.\\

\noindent Since the qubits $b_1, c_1, c_2$ are transferred through the phase flip noisy environment, according to the equation (7) the density matrix of the quantum channel is transformed into\\
$\epsilon^{PF} (\rho)= (1-\mu)^3 \bigg[ (|00000\rangle +|01011\rangle +|10100\rangle -|11111\rangle) \times (\langle00000| +\langle01011| +\langle10100| -\langle11111|)\bigg]\\
~~~~~~~~~~~~~~+2(1-\mu)^2 \mu \bigg[ (|00000\rangle -|01011\rangle +|10100\rangle +|11111\rangle) \times (\langle00000| -\langle01011| +\langle10100| +\langle11111|)\bigg]\\
~~~~~~~~~~~~~~+(1-\mu) \mu^2 \bigg[ (|00000\rangle +|01011\rangle +|10100\rangle -|11111\rangle) \times (\langle00000| +\langle01011| +\langle10100| -\langle11111|)\bigg]\\
~~~~~~~~~~~~~~+(1-\mu)^2 \mu \bigg[ (|00000\rangle +|01011\rangle -|10100\rangle +|11111\rangle) \times (\langle00000| +\langle01011| -\langle10100| +\langle11111|)\bigg]\\
~~~~~~~~~~~~~~+2(1-\mu) \mu^2 \bigg[ (|00000\rangle -|01011\rangle -|10100\rangle -|11111\rangle) \times (\langle00000| -\langle01011| -\langle10100| -\langle11111|)\bigg]\\
~~~~~~~~~~+\mu^3 \bigg[ (|00000\rangle +|01011\rangle -|10100\rangle +|11111\rangle) \times (\langle00000| +\langle01011| -\langle10100| +\langle11111|)\bigg].\\$

\noindent This can be written as\\
$\epsilon^{PF} (\rho)= \bigg((1-\mu)^3+(1-\mu) \mu^2 \bigg) \times \bigg[ (|00000\rangle +|01011\rangle +|10100\rangle -|11111\rangle) \times (\langle00000|\\
~~~~~~~~~~~~~~~ +\langle01011| +\langle10100| -\langle11111|)\bigg] +2(1-\mu)^2 \mu \bigg[ (|00000\rangle -|01011\rangle +|10100\rangle\\
~~~~~~~~~~~~~~~  +|11111\rangle) \times (\langle00000| -\langle01011| +\langle10100| +\langle11111|)\bigg] +\bigg((1-\mu)^2 \mu+\mu^3 \bigg) \\
~~~~~~~~~~~~~~~ \times \bigg[ (|00000\rangle+|01011\rangle -|10100\rangle +|11111\rangle) \times (\langle00000| +\langle01011| -\langle10100| \\
~~~~~~~~~~~~~~~+\langle11111|)\bigg]+2(1-\mu) \mu^2 \bigg[ (|00000\rangle -|01011\rangle -|10100\rangle -|11111\rangle) \times (\langle00000|\\
~~~~~~~~~~~~~~~-\langle01011| -\langle10100| -\langle11111|)\bigg]_{a_1a_2b_1c_1c_2}.\\$

\noindent Now Alice makes the measurement with her measuring basis given in (5) and communicates with Bob and Candy classically. In order to obtain the intended state, Bob and Candy make appropriate unitary operations on their respective qubits.\\

\noindent As an illustration, suppose Alice's measurement outcome is $|\xi_2\rangle$.  Therefore the density matrix of the final output states corresponding to Alice's measurement outcome $|\xi_2\rangle$ according to the formula (8) is given by\\
$\rho_{2}^{PF}= \bigg((1-\mu)^3+(1-\mu) \mu^2 \bigg) \times \bigg[ (\alpha \delta|011\rangle +\alpha \gamma|000\rangle +\beta \delta|111\rangle +\beta \gamma|100\rangle) \times (\alpha \delta\langle011|\\
~~~~~~~~~~~~~~~ +\alpha \gamma\langle000| +\beta \delta\langle111| +\beta \gamma\langle100|)\bigg] +2(1-\mu)^2 \mu \bigg[ (\alpha \delta|011\rangle -\alpha \gamma|000\rangle +\beta \delta|111\rangle\\
~~~~~~~~~~~~~~~  -\beta \gamma|100\rangle) \times (\alpha \delta\langle011| -\alpha \gamma\langle000| +\beta \delta\langle111| -\beta \gamma\langle100|)\bigg] +\bigg((1-\mu)^2 \mu+\mu^3 \bigg) \\
~~~~~~~~~~~~~~~ \times \bigg[ (\alpha \delta|011\rangle+\alpha \gamma|000\rangle -\beta \delta|111\rangle -\beta \gamma|100\rangle) \times (\alpha \delta\langle011| +\alpha \gamma\langle000| -\beta \delta\langle111| \\
~~~~~~~~~~~~~~~-\beta \gamma\langle100|)\bigg]+2(1-\mu) \mu^2 \bigg[ (\alpha \delta|011\rangle -\alpha \gamma|000\rangle -\beta \delta|111\rangle +\beta \gamma|100\rangle) \times (\alpha \delta\langle011|\\
~~~~~~~~~~~~~~~-\alpha \gamma\langle000| -\beta \delta\langle111| +\beta \gamma\langle100|)\bigg]_{b_1c_1c_2}.\\$

\noindent The fidelity of the output state with respect to Alice's measurement result $|\xi_2\rangle$ can be obtained according to the formula (10) as\\
$~~~~~~~F_2=\langle \Psi|\rho_2^{PF}|\Psi\rangle\\
~~~~~~~~~=\bigg((1-\mu)^3+(1-\mu) \mu^2 \bigg)+\bigg(2(1-\mu)^2 \mu (\delta^2 - \gamma^2)^2 \bigg)+\bigg((1-\mu)^2 \mu+\mu^3 \bigg)(\alpha^2 -\beta^2)^2\\
~~~~~~~~~~~~~+2(1-\mu) \mu^2 (\alpha^2 -\beta^2)^2 (\delta^2 - \gamma^2)^2.$
\begin{figure}[t]
\begin{subfigure}{0.5\textwidth}
\includegraphics[width=0.9\linewidth, height=5cm]{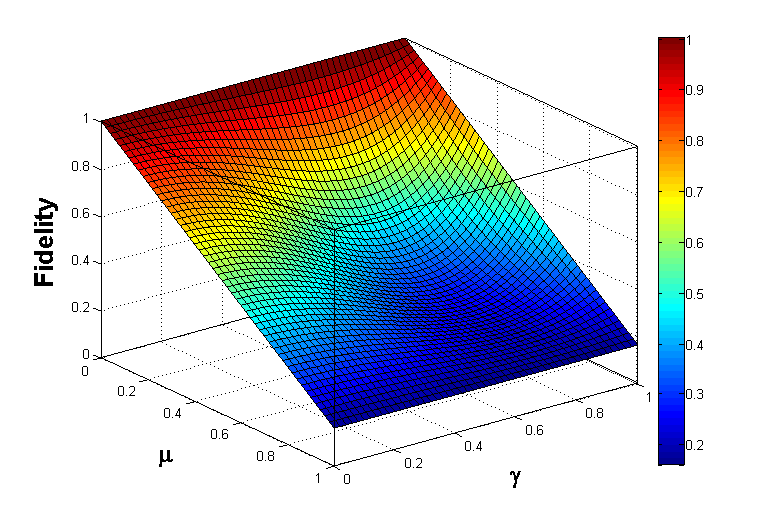} 
\caption{ }
\label{}
\end{subfigure}
\begin{subfigure}{0.5\textwidth}
\includegraphics[width=0.9\linewidth, height=5cm]{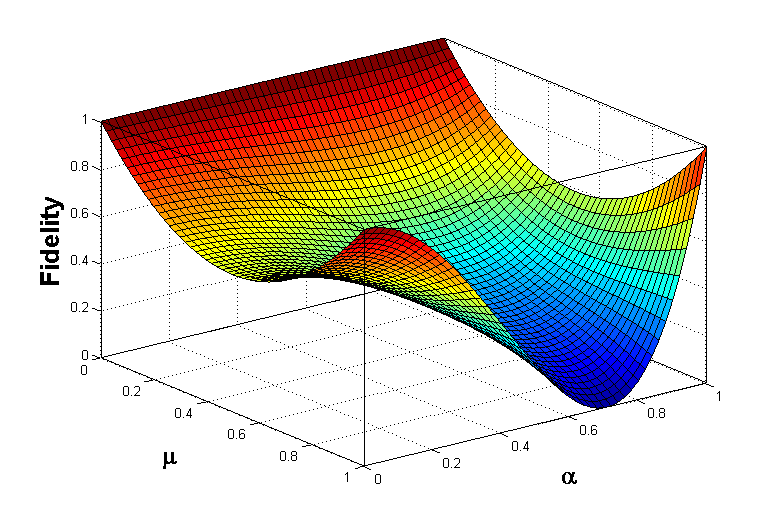} 
\caption{ }
\label{}
\end{subfigure}
\begin{subfigure}{0.5\textwidth}
\includegraphics[width=0.9\linewidth, height=5cm]{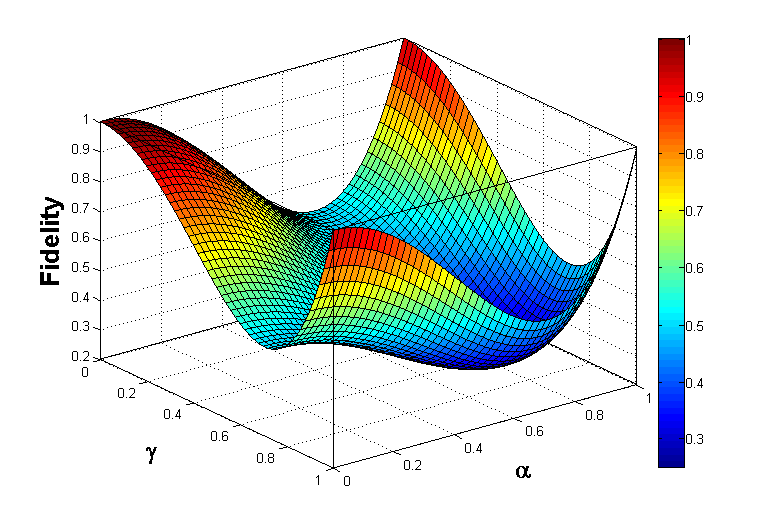} 
\caption{ }
\label{}
\end{subfigure}
\begin{subfigure}{0.5\textwidth}
\includegraphics[width=0.9\linewidth, height=5cm]{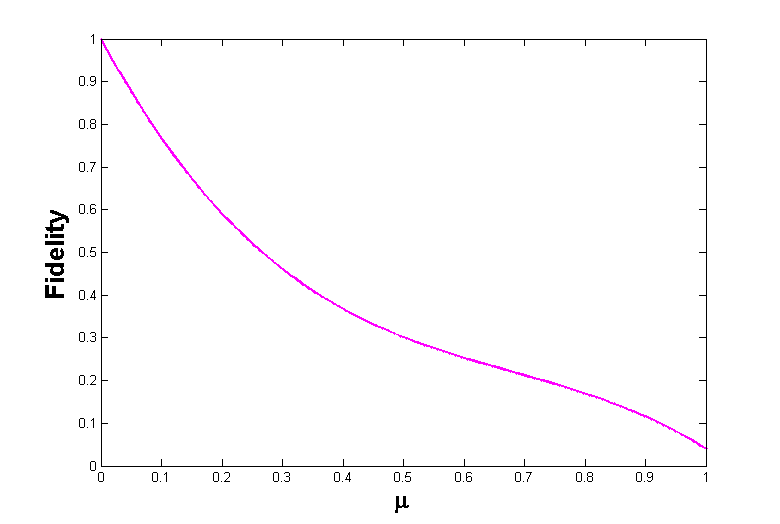} 
\caption{ }
\label{}
\end{subfigure}
\caption{(a): Variation of fidelity $F_2$  with $\gamma$ and $\mu$ when $ |S_1\rangle=(\sqrt{0.3}|0\rangle+\sqrt{0.7}|1\rangle)$ (b): Variation of fidelity $F_2$  with $\alpha$ and $\mu$ when  $|S_2\rangle=(\sqrt{0.3}|00\rangle+\sqrt{0.7}|11\rangle)$ (c): Variation of fidelity $F_2$  with $\alpha$ and  $\gamma$ when  $\nu=0.5$ (d) Variation of fidelity $F_2$  with $\mu$ when $ |S_1\rangle=(\sqrt{0.4}|0\rangle+\sqrt{0.6}|1\rangle)$ and $|S_2\rangle=(\sqrt{0.3}|00\rangle+\sqrt{0.7}|01\rangle)$. }
\label{fig:2}
\end{figure}
\noindent \subsection{\textbf{ Bit-flip noisy environment  }}
The Kraus operators for the bit-flip noisy environment are given by\\
$$X_0=\left( 
\begin{array}{rr}
\sqrt{1-\nu} & 0 \\ 
0 & \sqrt{1-\nu}%
\end{array}%
\right)=\sqrt{1-\nu} I~~~~~~~~~~X_1=\left( 
\begin{array}{rr}
0 & \sqrt{\nu}\\ 
\sqrt{\nu} & 0%
\end{array}%
\right)=\sqrt{\nu} \sigma_x,$$
where $\nu \in [0,1]$ indicates the noise intensity parameter in bit-flip noisy environment.\\

\noindent According to the formula given in (7), the density matrix of the transformed entangled channel (as the qubits $b_1, c_1, c_2$ pass through a bit-flip noisy environment channel) is given by\\
$\epsilon^{BF} (\rho)=(1-\nu)^{3} \bigg[ (|00000\rangle +|01011\rangle +|10100\rangle -|11111\rangle) \times (\langle00000| +\langle01011| +\langle10100|-\langle11111|)\bigg]\\
~~~~~~~~~~~~+(1-\nu)^2 \nu \bigg[ (|00001\rangle +|01010\rangle +|10101\rangle -|11110\rangle) \times (\langle00001| +\langle01010| +\langle10101|-\langle11110|)\bigg]\\
~~~~~~~~~~~~+(1-\nu)^2 \nu \bigg[ (|00010\rangle +|01001\rangle +|10110\rangle -|11101\rangle) \times (\langle00010| +\langle01001| +\langle10110|-\langle11101|)\bigg]\\
~~~~~~~~~~~~+\nu^2 (1-\nu) \bigg[ (|00011\rangle +|01000\rangle +|10111\rangle -|11100\rangle) \times (\langle00011| +\langle01000| +\langle10111|-\langle11100|)\bigg]\\
~~~~~~~~~~~~+(1-\nu)^2 \nu \bigg[ (|00100\rangle +|01111\rangle +|10000\rangle -|11011\rangle) \times (\langle00100| +\langle01111| +\langle10000|-\langle11011|)\bigg]\\
~~~~~~~~~~~~+\nu^2 (1-\nu) \bigg[ (|00101\rangle +|01110\rangle +|10001\rangle -|11010\rangle) \times (\langle00101| +\langle01110| +\langle10001|-\langle11010|)\bigg]\\
~~~~~~~~~~~~+\nu^2 (1-\nu) \bigg[ (|00110\rangle +|01101\rangle +|10010\rangle -|11001\rangle) \times (\langle00110| +\langle01101| +\langle10010|-\langle11001|)\bigg]\\
~~~~~~~~~~~~+\nu^{3} \bigg[ (|00111\rangle +|01100\rangle +|10011\rangle -|11000\rangle) \times (\langle00111| +\langle01100| +\langle10011|-\langle11000|)\bigg].\\$

\noindent Now Alice performs the measurement on her two qubits $a_1, a_2$ with the basis given in (5) and sends the outcome results classically to Bob and Candy. In order to obtain the intended state, Bob and Candy make appropriate unitary operations on their respective qubits.\\

\noindent As an illustration, suppose Alice's measurement outcome is $|\xi_2\rangle$. Then Bob and Candy apply an appropriate unitary operations $(\sigma_z)_{b_1}$ and $(\sigma_x)_{c_1} \otimes (\sigma_x\sigma_z)_{c_2}$ on their respective qubits. Therefore, according to the formula given in (8) the reduced density matrix of the final output state corresponding to Alice's measurement result $|\xi_2\rangle$ is given by\\
$\rho_2^{BF} (\rho)=(1-\nu)^{3} \bigg[ (\alpha \delta|011\rangle +\alpha \gamma|000\rangle +\beta \delta|111\rangle +\beta \gamma|100\rangle) \times (\alpha \delta\langle011| +\alpha \gamma\langle000| +\beta \delta\langle111|\\
~~~~~~~~~~~~+\beta \gamma\langle100|)\bigg]+(1-\nu)^2 \nu \bigg[ (-\alpha \delta|010\rangle -\alpha \gamma|001\rangle -\beta \delta|110\rangle -\beta \gamma|101\rangle) \times (-\alpha \delta\langle010| -\alpha \gamma\langle001|\\
~~~~~~~~~~~~ -\beta \delta\langle110|-\beta \gamma\langle101|)\bigg]+(1-\nu)^2 \nu \bigg[ (\alpha \delta|001\rangle +\alpha \gamma|010\rangle +\beta \delta|101\rangle +\beta \gamma|110\rangle) \times (\alpha \delta\langle001|\\
~~~~~~~~~~~~ +\alpha \gamma\langle010| +\beta \delta\langle101|+\beta \gamma\langle110|)\bigg]+\nu^2 (1-\nu) \bigg[ (-\alpha \delta|000\rangle -\alpha \gamma|011\rangle -\beta \delta|100\rangle -\beta \gamma|111\rangle) \\
~~~~~~~~~~~~\times (-\alpha \delta\langle000| -\alpha \gamma\langle011| -beta \delta\langle100|- \beta \gamma\langle111|)\bigg]+(1-\nu)^2 \nu \bigg[ (-\alpha \delta|111\rangle -\alpha \gamma|100\rangle -\beta \delta|011\rangle \\
~~~~~~~~~~~~-\beta \gamma|000\rangle) \times (-\alpha \delta\langle111| -\alpha \gamma\langle100| -\beta \delta\langle011|-\beta \gamma\langle000|)\bigg]+\nu^2 (1-\nu) \bigg[ (\alpha \delta|110\rangle +\alpha \gamma|101\rangle \\
~~~~~~~~~~~~+\beta \delta|010\rangle +\beta \gamma|001\rangle) \times (\alpha \delta\langle110| +\alpha \gamma\langle101| +\beta \delta\langle010|+\beta \gamma\langle001|)\bigg]+\nu^2 (1-\nu) \bigg[ (-\alpha \delta|101\rangle\\
~~~~~~~~~~~~-\alpha \gamma|110\rangle -\beta \delta|001\rangle -\beta \gamma|010\rangle) \times (-\alpha \delta\langle101| -\alpha \gamma\langle110| -\beta \delta\langle001|-\beta \gamma\langle010|)\bigg]+\nu^{3}\\
~~~~~~~~~~~~ \times\bigg[ (\alpha \delta|100\rangle +\alpha \gamma|111\rangle +\beta \delta|000\rangle +\beta \gamma|011\rangle) \times (\alpha \delta\langle100| +\alpha \gamma\langle111| +\beta \delta\langle1000|+\beta \gamma\langle011|)\bigg].\\$

\noindent The fidelity of the final output state can be obtained according to the formula (10) as\\
$~~~~~~~F_3=\langle \Psi|\rho_2^{BF}|\Psi\rangle\\
~~~~~~~~~=(1-\nu)^{3}+4\nu^2 (1-\nu)\gamma^2 \delta^2 +4(1-\nu)^2 \nu\alpha^2 \beta^2+16\nu^{3}\alpha^2 \beta^2\gamma^2 \delta^2.$

\begin{figure}[t]
\begin{subfigure}{0.5\textwidth}
\includegraphics[width=0.9\linewidth, height=5cm]{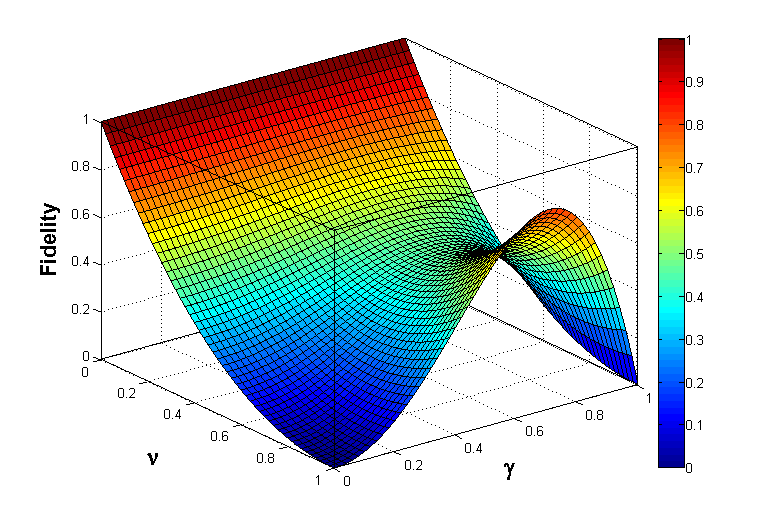} 
\caption{ }
\label{}
\end{subfigure}
\begin{subfigure}{0.5\textwidth}
\includegraphics[width=0.9\linewidth, height=5cm]{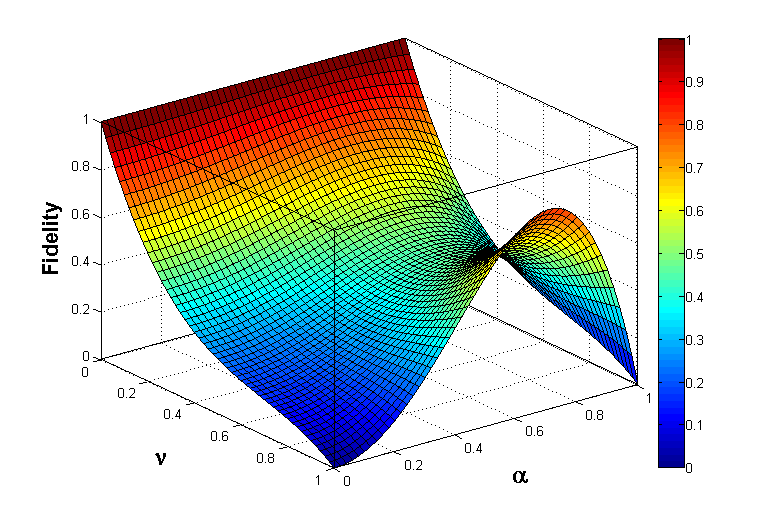} 
\caption{ }
\label{}
\end{subfigure}
\begin{subfigure}{0.5\textwidth}
\includegraphics[width=0.9\linewidth, height=5cm]{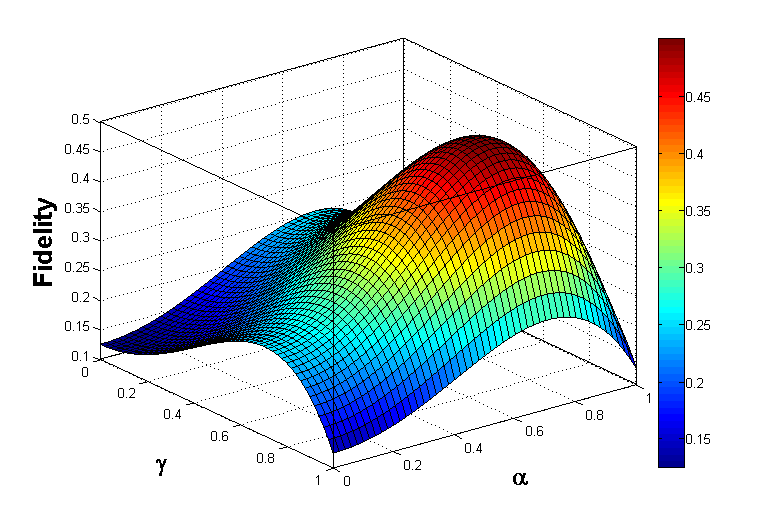} 
\caption{ }
\label{}
\end{subfigure}
\begin{subfigure}{0.5\textwidth}
\includegraphics[width=0.9\linewidth, height=5cm]{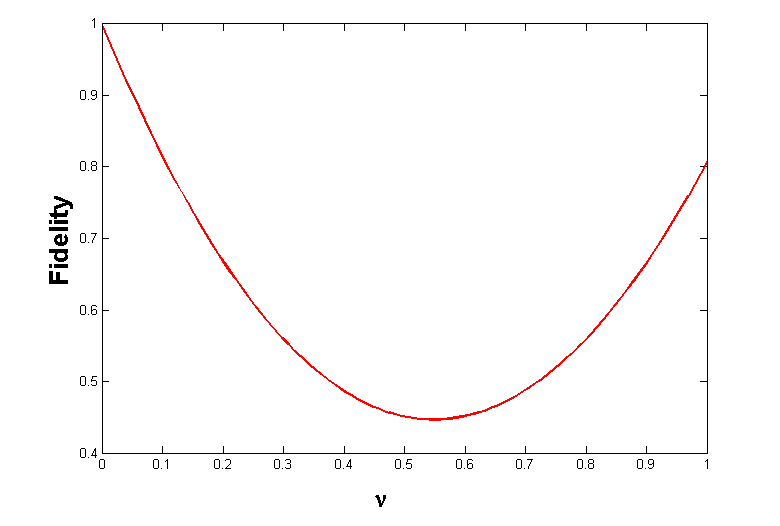} 
\caption{ }
\label{}
\end{subfigure}
\caption{(a): Variation of fidelity $F_3$  with $\gamma$ and $\mu$ when $ |S_1\rangle=(\sqrt{0.3}|0\rangle+\sqrt{0.7}|1\rangle)$ (b): Variation of fidelity $F_3$  with $\alpha$ and $\mu$ when  $|S_2\rangle=(\sqrt{0.3}|00\rangle+\sqrt{0.7}|11\rangle)$  (c): Variation of fidelity $F_3$  with $\alpha$ and  $\gamma$ when  $\nu=0.5$ (d) Variation of fidelity $F_3$  with $\mu$ when $ |S_1\rangle=(\sqrt{0.4}|0\rangle+\sqrt{0.6}|1\rangle)$ and $|S_2\rangle=(\sqrt{0.3}|00\rangle+\sqrt{0.7}|01\rangle)$. }
\label{fig:2}
\end{figure}

\section{\textbf{Discussion and Conclusion}}
The speciality of the noisy environment considered here is that all the qubits which are transmitted experience noise individually, that is, there is no correlation regarding the noisy environment between two qubits even when they are sent to the same party in the same prevalent noisy environment. This is reflected in the expression of equation (7) where the `$\sum$' is over the indices $l, m$ and $n$. It is known that there are other types of noises than those considered here which can be modeled using the Kraus operator. There is no intrinsic reason for our consideration of only these three types of representative noises. Also it is known that the effect of noise can be minimized by executing weak and  reversal measurements. In the present context we have not considered the effects of such noise-minimizing mechanisms. That would have warranted  a separate work by themselves. Nonetheless, these are important aspects of communication that can be taken up in future works.\\

\noindent It is interesting to see the variation of fidelity with the variation of other parameters given in Figures 3, Figure 4 and Figure 5. The common feature is that the fidelity in all cases go to unit value as the noise parameter tends to zero. This is as should be expected since with the noise parameter going to zero we approach the perfect remote state preparation protocol described in section $2$  where the value of fidelity is $1$.\\

\noindent Further the communication scheme presented here is a multi-tasking protocol where more than one quantum information tasks are performed with the help of a single entanglement resource. With the increase in the complex nature of the jobs to be performed there is an increased necessity of entanglement with more number of qubits serving the communication scheme. Multi-partite entanglement is difficult to produce. Here our necessity is of a five-qubit entanglement. Preparation of the entanglement resource is designed in section $3$ and is run on IBM platform.\\

\noindent\textbf{Acknowledgement} This work is supported by the Indian Institute of Engineering Science and Technology, Shibpur.\\

\noindent\textbf{Data availability statement} Our manuscript has no associated data.\\

\noindent\textbf{Conflict of interest} On behalf of all authors, the corresponding author states that there is no conflict of interest.\\

\end{document}